\documentclass[aps, showkeys, reprint, onecolumn, superscriptaddress, notitlepage]{revtex4-1}

\usepackage[english]{babel}
\usepackage[utf8]{inputenc}
\usepackage{amsmath}
\usepackage{amsfonts}
\usepackage{amssymb}
\usepackage{graphicx}
\usepackage{array}
\usepackage{epsfig}

\newcommand{\diff}{\text{d}}

\usepackage{xcolor}
\definecolor{mycolor3}{rgb}{0.63500,0.07800,0.18400}	

\begin{document}

\title{Experimental self-generation of axisymmetric internal wave super-harmonics}
\date{\today}

\author{S. Boury}
\affiliation{Courant Institute of Mathematical Sciences, New York University, New York, NY 10012, USA}
\affiliation{Univ Lyon, ENS de Lyon, Univ Claude Bernard, CNRS, Laboratoire de Physique, F-69342 Lyon, France}
\affiliation{Department of Mechanical Engineering, Massachusetts Institute of Technology, Cambridge, MA 02139, USA}
\author{T. Peacock}
\affiliation{Department of Mechanical Engineering, Massachusetts Institute of Technology, Cambridge, MA 02139, USA}
\author{P. Odier}
\affiliation{Univ Lyon, ENS de Lyon, Univ Claude Bernard, CNRS, Laboratoire de Physique, F-69342 Lyon, France}

	\begin{abstract}
	
	In this paper, we present an experimental study of weakly non-linear interaction of axisymmetric internal gravity waves in a resonant cavity, supported by theoretical considerations. Contrary to plane waves in Cartesian coordinates, for which self-interacting terms are null in a linear stratifiation, the non-linear self-interaction of an internal wave mode in axisymmetric geometry is found to be efficient at producing super-harmonics, i.e. waves whose frequencies are integer multiples of the excitation frequency. Due to the range of frequencies tested in our experiments, the first harmonic frequency is below the cut-off imposed by the stratification so the lowest harmonic created can always propagate. The study shows that the super-harmonic wave field is a sum of standing waves satisfying both the dispersion relation for internal waves and the boundary conditions imposed by the cavity walls, while conserving the axisymmetry.	
	\end{abstract}
	\maketitle
	
	\section{Introduction}
	
	In a recent review paper, MacKinnon~\textit{et al.}~\cite{mackinnon2017BAMS} discussed the mechanisms and implications of the dissipation of internal wave energy in the oceans, revisiting the $2\pm0.6\mathrm{~TW}$ estimate of turbulent dissipation caused by tidal flow over topographies, low-frequency lee waves, and near-inertial waves produced by wind forcing \cite{waterhouse2014, kunze2017, baker2020}. In these various scenarios, through non-linear interactions, internal wave breaking can transfer energy to smaller scales at which it can be dissipated more efficiently. Evidenced by the experimental work of Joubaud \textit{et al.} \cite{joubaud2012}, Triadic Resonant Instability (TRI) has been proven relevant for wave breaking and for triggering non-linear sub-harmonic resonant cascades of highly energetic waves \cite{brouzet2016a}. A very similar self-interaction mechanism has been predicted and tested numerically as a means to generate sub- and super-harmonic waves, at different scales, through the interaction with either topography or stratification \cite{wunsch2015, sutherland2016, varma2017}. Super-harmonics have also been proposed as a way to transfer energy to other scales and even as a preliminary step before producing small scale waves~\cite{wagner2016}.
	
	To date, non-linear interactions have been extensively investigated in two-dimensional Cartesian geometry (for instance in \cite{joubaud2012, wunsch2015, brouzet2016a, sutherland2016, varma2017}) but three-dimensional studies remain marginal (for example, see the recent work of \cite{shmakova2018}). Arguably relevant to geophysics, as they mimic the geometry of a localised wave source in unconfined domains, axisymmetric wave fields are still challenging to study when dealing with linear and non-linear phenomena. Classical generation mechanisms, such as vertically oscillating spheres \cite{mowbray1967a, ermanyuk2011}, are challenging for the study of $3$D non-linearities as they usually generate quite spatially localized 3D wave fields. Instead, Maurer \textit{et al.}'s axisymmetric wave generator \cite{maurer2017}, built upon the technology of Gostiaux \textit{et al.}'s planar generator \cite{gostiaux2006}, has proven capable of exciting pure axisymmetric wave fields shaped as Bessel functions \cite{boury2018}, which can enable more practical studies of non-linear effects.
	
	In non-linear stratifications, ubiquituous in geophysical flows, wave-wave interaction has been demonstrated as being capable of producing higher frequency harmonics, either through the forced interaction of two different waves \cite{husseini2019}, or through the self-interaction of the wave itself \cite{sutherland2016, baker2020} (note that, in what follows, we will use the term self-interaction for the interaction of a single monochromatic wave with itself). Interestingly, for non-rotating flows, $2$D Cartesian self-interacting terms are null in linear stratifications and the non-uniform buoyancy frequency is therefore a condition for the existence of super-harmonics. Extending these works to $3$D wave fields, however, remains challenging, and poorly investigated experimentally. Closed domains, such as the resonant cavity described in \citep{boury2018}, are capable of sustaining enhanced amplitude wave fields and non-linear interactions both in linear and non-linear stratifications \cite{boury2018,boury2019}, and could possibly lead to TRI or super-harmonic generation. Through the use of an oscillating sphere and the study of its radiated conical wave field, Ermanyuk~\textit{et al.} \cite{ermanyuk2011} have shown that the first harmonic created by non-linear interaction is likely to produce a non-trivial structure with a symmetry breaking, e.g. dipolar or quadripolar. In their study, however, the radiated wave field cannot be decomposed over a single Bessel function and the different components of the wave field, which can be extracted through a Fourier-Hankel transform \cite{maurer2017}, can interact together \textit{via} the non-linear terms of the wave equation. In the present work, we consider a purely axisymmetric wave field shaped as a radial standing wave, called a radial mode, and discuss what is, to our knowledge, the first experimental observation of super-harmonic generation through self-interaction of axisymmetric gravity waves.
	
	The paper is organised as follows. First, we derive the theoretical framework for linear and weakly non-linear internal wave propagation. This section also discusses the wave resonator and modal selection induced by the particular closed geometry, a cylindrical cavity, giving relevant insights on the observed phenomenon. After presenting the experimental apparatus in section $3$, the results are described in section $4$ and the performed analysis allows for a complete description of the created harmonics. Finally, our conclusions and discussion are presented in section $5$.
	
	\section{Theory}
	
		\subsection{Governing Equations}
		
		In a cylindrical framework ($\mathbf{e_r}$, $\mathbf{e_\theta}$, $\mathbf{e_z}$), $\mathbf{e_z}$ being the ascendent vertical, small amplitude inertia gravity waves in an inviscid fluid with a constant background stratification satisfy the following equations under the Boussinesq approximation
		\begin{eqnarray}
			\rho_0 \left(\dfrac{\partial \mathbf{v}}{\partial t} + \left(\mathbf{v}\cdot\mathbf{\nabla}\right) \mathbf{v}\right) &=& - \mathbf{\nabla} p - (\rho - \bar{\rho}) g \mathbf{e_z},\label{eq:IW1}\\
			\dfrac{\partial \rho}{\partial t} + \left(\mathbf{v}\cdot \mathbf{\nabla}\right) \rho &=& 0,\label{eq:IW2}\\
			\mathbf{\nabla} \cdot \mathbf{v} &=& 0\label{eq:IW3},
		\end{eqnarray}
		where $\mathbf{v} = (v_r,~v_\theta,~v_z)$, $p$, $\rho$, and $\bar{\rho}$ are the velocity, pressure, density, and background density fields, respectively. Setting $\rho_0$ a reference density, the buoyancy field $b$ and frequency $N$ can be introduced as
		\begin{equation}	
			b = -\frac{g}{\rho_0} \rho' \mathrm{~~~~~~~and~~~~~~~} N^2 = -\frac{g}{\rho_0} \dfrac{\partial \bar{\rho}}{\partial z},\label{eq:IW4}
		\end{equation}
		where $\rho' = \rho - \bar{\rho}$ is the perturbation to the density field.
		
		In an axisymmetric geometry, the velocity and the buoyancy fields are $\theta$-independent and the vertical and radial velocities are therefore given by derivatives of a stream function $\psi$ as
		\begin{equation}
			v_r = - \frac{1}{r}\dfrac{\partial (r \psi)}{\partial z} \mathrm{~~~~~~~and~~~~~~~} v_z =\frac{1}{r}\dfrac{\partial (r \psi)}{\partial r}.\label{eq:IW5}
		\end{equation}
		Using this formulation, the system of equations~\eqref{eq:IW1},~\eqref{eq:IW2}, and~\eqref{eq:IW3} can be written as coupled equations in $\psi$, $v_\theta$, and $b$,
		\begin{eqnarray}
		\partial_t \Delta_h \psi + \mathcal{J}^\odot \left(r\psi, \frac{\Delta_h \psi}{r}\right) &=& - \frac{2}{r} v_\theta \partial_z v_\theta + \partial_r b, \label{eq:IW6}\\
		\partial_t v_\theta + \frac{1}{r^2} \mathcal{J}^\odot(r\psi, r v_\theta) &=& 0, \label{eq:IW7}\\
		\partial_t b + \frac{1}{r} \mathcal{J}^\odot(r\psi, b) &=& - N^2 \frac{1}{r} \partial_r (r\psi),\label{eq:IW8}
	\end{eqnarray}
	where the truncated Laplacian $\Delta_h \psi$ is defined by
	\begin{equation}
		\Delta_h \psi = \dfrac{\partial^2 \psi}{\partial z^2} + \dfrac{\partial}{\partial r} \left( \frac{1}{r} \dfrac{\partial (r\psi)}{\partial r} \right) = \dfrac{\partial^2 \psi}{\partial z^2} + \dfrac{\partial^2 \psi}{\partial r^2} + \frac{1}{r}\dfrac{\partial \psi}{\partial r} -\frac{\psi}{r^2},\label{eq:IW9}
	\end{equation}
	and the cylindrical Jacobian $\mathcal{J}^\odot$ of two functions $f$ and $g$ is given by
	\begin{equation}
		\mathcal{J}^\odot (f,g) = \dfrac{\partial f}{\partial r}\dfrac{\partial g}{\partial z} - \dfrac{\partial g}{\partial r}\dfrac{\partial f}{\partial z}.\label{eq:IW10}
	\end{equation}
						
		\subsection{Solution of the Linear Problem}
		
			The system of equations \eqref{eq:IW6}, \eqref{eq:IW7}, and \eqref{eq:IW8} can be linearised by setting all Jacobians and the $v_\theta\partial_z v_\theta$ terms equal to zero, leading to the linear problem
			\begin{eqnarray}
				\partial_t \Delta_h \psi &=& \partial_r b, \label{eq:IW11}\\
				\partial_t v_\theta &=& 0, \label{eq:IW12}\\
				\partial_t b &=& - N^2 \frac{1}{r} \partial_r (r\psi).\label{eq:IW13}
			\end{eqnarray}
			Equation \eqref{eq:IW12} is self-consistent, and yields an orthoradial velocity that does not evolve in time. On the other hand, cross-derivatives of equations \eqref{eq:IW11} and \eqref{eq:IW13} give a linear equation in $\psi$
			\begin{equation}
				\mathcal{L}[\psi] = 0,\label{eq:IW14}
			\end{equation}
			where $\mathcal{L}$ is an operator defined as
			\begin{equation}
				\mathcal{L}[\psi] = \partial^2_t \Delta_h \psi + N^2\partial_r\left(\frac{1}{r} \partial_r (r\psi) \right).\label{eq:IW15}
			\end{equation}
			As discussed in Maurer~\textit{et al.}~\cite{maurer2017} and in Boury~\textit{et al.}~\citep{boury2018}, the linear solution of equations \eqref{eq:IW12} and \eqref{eq:IW14} that fulfills the boundary conditions imposed by an axisymmetric forcing at the surface and zero velocities normal to the other sides, with no orthoradial velocity, is
			\begin{eqnarray}
				\psi (r,z,t) &=& \psi_0 J_1 (l r) \cos (\omega t - m z),\label{eq:IW16}\\
				v_\theta (r,z,t) &=& 0,\label{eq:IW17}\\
				b (r,z,t) &=& -\frac{N^2 l}{\omega} \psi_0 J_0 (l r) \sin (\omega t - m z),\label{eq:IW18}
			\end{eqnarray}
			with $\omega$ the frequency of the wave field, $l$ and $m$ the radial and the vertical wave numbers, and $\psi_0$ a constant amplitude. The radial structure of the wave field is given by Bessel functions of zeroth and first orders, namely $J_0$ and $J_1$. With the proposed solution \eqref{eq:IW16} for $\psi$, the linear operator \eqref{eq:IW15} gives the dispersion relation for gravity waves
			\begin{equation}
				(l^2 + m^2) \omega^2 = N^2 l^2.\label{eq:IW19}
			\end{equation}
					
		\subsection{Non-Linear Problem}
		
			Back to equations \eqref{eq:IW6}, \eqref{eq:IW7}, and \eqref{eq:IW8}, the same cross-derivation that led to the linear system can be used to derive the fully non-linear equation
			\begin{equation}
				\mathcal{L}[\psi] = \mathcal{N}[\psi,v_\theta,b],\label{eq:IW20}
			\end{equation}
			where $\mathcal{L}$ is still defined by equation \eqref{eq:IW15} and $\mathcal{N}$ is the non-linear operator
			\begin{equation}
				\mathcal{N}[\psi,v_\theta,b] = -\partial_t \mathcal{J}^\odot \left(r\psi, \frac{\Delta_h \psi}{r}\right) - \partial_r \left( \frac{1}{r} \mathcal{J}^\odot(r\psi, b) \right) - \frac{1}{r}\partial_t \partial_z v_\theta^2.\label{eq:IW21}
			\end{equation}
			As discussed in the appendix of \cite{boury2019}, when the velocity amplitudes become too large, these non-linear terms cannot be neglected. If we consider the linear solution, we note that $v_\theta = 0$ so $\mathcal{N}[\psi,v_\theta,b] = \mathcal{N}[\psi,b]$. The non-linear right-hand side of equation \eqref{eq:IW20} therefore becomes only dependent on two Jacobians that can be evaluated. Using that \cite{nist2010}
			\begin{equation}
				\dfrac{\partial (J_0 (l r))}{\partial r} = - l J_1 (l r), \mathrm{~~~~~~~}\dfrac{\partial (r J_1 (l r))}{\partial r} = l r J_0 (l r), \mathrm{~~~~~~~and~~~~~~~}\dfrac{\partial (r^{-1} J_1 (l r))}{\partial r} = - l r^{-1} J_2 (l r),\label{eq:IW22}
			\end{equation}
			and $\psi$ and $b$ as derived in equations~\eqref{eq:IW16} and~\eqref{eq:IW18}, the non-linear terms involved in $\mathcal{N}[\psi,b]$ become
			\begin{eqnarray}
				\partial_t \mathcal{J}^\odot \left(r\psi, \frac{\Delta_h \psi}{r}\right) &=& C \frac{\left[J_1 (l r)\right]^2}{r} \cos (2 \omega t - 2 m z),\label{eq:IW23}\\
				\partial_r \left( \frac{1}{r} \mathcal{J}^\odot(r\psi, b) \right) &=& C\left[ \cos^2 (\omega t - m z) J_0(lr) \partial_r J_0(lr) + \sin^2 (\omega t -m z) J_1 (l r) \partial_r J_1 (l r) \right],\label{eq:IW24}
			\end{eqnarray}
			with
			\begin{equation}
				C = 2 \omega (l^2 + m^2) m \psi_0^2.\label{eq:IW25}
			\end{equation}
			As a result, the non-linear right-hand side of equation~\eqref{eq:IW20} is
			\begin{equation}
				\mathcal{N}[\psi,b] = C J_1 (l r) \left[ J_2 (l r) \sin^2 (\omega t - m z) - J_1 (l r) \cos^2 (\omega t - m z) \right] \neq 0\label{eq:IW26}
			\end{equation}
			
			We can go a step further if, in the general case, inspired by the work of Thorpe~\cite{thorpe1966}, we write $\psi$ as a sum over possible solutions of various frequencies $\omega_j$ as
			\begin{equation}
				\psi(r,z,t) = \sum_j \chi_j (r,z) \cos(\omega_j t),\label{eq:IW27}
			\end{equation}
			very much like a discrete Fourier Transform, with $\chi_j$, $j\in\mathbb{Z}$, a spatial function. The linear part of equation \eqref{eq:IW20} is therefore
			\begin{equation}
				\mathcal{L}[\psi] = \sum_j \left[(N^2 - \omega_j^2)\dfrac{\partial}{\partial_r}\left( \frac{1}{r}\dfrac{\partial (r \chi_j)}{\partial_r}\right) - \dfrac{\partial^2 \chi_j}{\partial z^2}\right] \cos(\omega_j t).\label{eq:IW28}
			\end{equation}
			As the non-linear terms are only second order products of the stream function, we assume the following development
			\begin{equation}
				\mathcal{N}[\psi] = \sum_{g,h} C_{g,h} (r,z) \cos((\omega_g+\omega_h)t), \label{eq:IW29}
			\end{equation}
			where $\omega_g$ and $\omega_h$ are frequencies, and $C_{g,h}$ is a spatial fonction. Projection of equation \eqref{eq:IW21} over frequencies leads to
			\begin{equation}
				\forall j \in \mathbb{Z},~(N^2 - \omega_j^2)\dfrac{\partial}{\partial_r}\left( \frac{1}{r}\dfrac{\partial (r \chi_j)}{\partial_r}\right) - \dfrac{\partial^2 \chi_j}{\partial z^2} = C_j,\label{eq:IW30}
			\end{equation}
			where $C_j$ is defined as the function $C_{g,h}$ where $g$ and $h$ verify $\omega_j = \omega_g + \omega_h$. The $C_j$ functions act as forcing terms produced by non-linear wave-wave interactions. The impact of this forcing term can be further explored theoretically by using Green functions (see appendix) and a Taylor expansion of the stream function, or with Direct Numerical Simulations, which is beyond the scope of this study.
					
		\subsection{Wave Resonator and Mode Selection}
		
		Cylindrical cavities have the ability to produce enhanced modal wave fields, likely to trigger instabilities at high frequencies~\cite{boury2018}. In such a confined configuration, radial and vertical modes can be excited if they satisfy the boundary condition, namely a zero orthogonal velocity at the boundaries. Hence, a mode can be described by the stream function
		\begin{equation}
			\psi(r,z,t) = \psi^0 J_1(l r) \sin (m z) \cos(\omega t).\label{eq:IW31}
		\end{equation}
		As we will see next, the radial confinement sets the possible values of $l$, and so does the vertical confinement for the values of $m$.
		
		\subsubsection{Radial Confinement}
		
		The cylinder imposes a radial confinement within a radius $R$. As such, modes of radial wave number $l_p, ~p\in \mathbb{N}^*$, are selected with the boundary condition stating that the radial velocity is zero at the boundary
		\begin{equation}
			v_r (r = R, z, t) = \left(-\frac{1}{r}\dfrac{\partial (r \psi)}{\partial z} \right)_{r=R} = 0,\label{eq:IW32}
		\end{equation}
		equivalent to
		\begin{equation}
			 J_1(l_p R) = 0.\label{eq:IW33}
		\end{equation}
		If $j_{1,p}$ is the $p$\textsuperscript{th} zero of the $1$\textsuperscript{st} order Bessel function $J_1$, then the values of $l_p$ are given by
		\begin{equation}
			\forall p \in \mathbb{N}^*,~l_p = \frac{j_{1,p}}{R}.\label{eq:IW34}
		\end{equation}
		Table~\ref{tab:lp} presents the first zeros of the $J_1$ Bessel function, extracted from~\cite{beattie1958}, and the corresponding radial wave numbers for a cylindrical confinement of radius $R = 20\mathrm{~cm}$. An experimental visualisation of these radial modes for $p=1$, $2$, and $3$, can be found in Boury~\textit{et al.}~\cite{boury2018}.
		\begin{table}[!htbp]
			\centering
			\begin{tabular}{c  c c c c c}
				\hline 
				$p$ & $1$ & $2$ & $3$ & $4$ & $5$ \\ 
				\hline \hline
				$j_{1,p}$ & $3.83$ & $7.02$ & $10.17$ & $13.32$ & $16.47$ \\ 
				$l_p$ & $19.15$ & $35.10$ & $50.85$ & $66.60$ & $82.35$ \\ 
				\hline
			\end{tabular} 		
			\caption{First zeros of the $J_1$ Bessel function and associated radial wave number for $R=20\mathrm{~cm}$.}
			\label{tab:lp}
		\end{table}
		
		\subsubsection{Vertical Confinement}
		
		The upper and lower boundaries, set by the bottom of the tank and the wave generator, respectively, impose a vertical confinement (cavity) over a given height $L$. The vertical modes are found by stating that, at $z=0$ (when neglecting the low amplitude generator motion) and at $z=-L$, the vertical velocity is zero, i.e.
		\begin{equation}
			v_z (z=0) = \left(\frac{1}{r}\dfrac{\partial (r\psi)}{\partial r}\right) _ {z=0} = 0 \mathrm{~~~and~~~} v_z (z=-L) = \left(\frac{1}{r}\dfrac{\partial (r\psi)}{\partial r}\right) _ {z=-L} = 0,\label{eq:IW35}
		\end{equation}
		which justifies that the vertical dependence of $\psi$ is described by a sine function, and also means that
		\begin{equation}
			\sin(m L) = 0.\label{eq:IW36}
		\end{equation}
		As a consequence, vertical modes should have a half-integer number of wave lengths in the cavity. It follows that the vertical wave number $m_q,~q\in\mathbb{N}^*$, is given by
		\begin{equation}
			\forall q \in \mathbb{N}^*,~m_q = \frac{\pi q}{L}.\label{eq:IW37}
		\end{equation}
		Table~\ref{tab:mq} gives the smallest values of $m_q$ that can be found in a resonant cavity of height $L=60\mathrm{~cm}$.
		\begin{table}[!htbp]
			\centering
			\begin{tabular}{c c c c c c c c c c}
				\hline 
				$q$ & $1$ & $2$ & $3$ & $4$ & $5$ & $6$ & $7$ & $8$ & $9$\\ 
				\hline \hline
				$m_q$ & $5.23$ & $10.47$ & $15.70$ & $20.93$ & $26.17$ & $31.40$ & $36.63$ & $41.87$ & $47.1$ \\ 
				\hline 
			\end{tabular} 		
			\caption{Lowest vertical wave numbers $m_q$ for $L=60\mathrm{~cm}$.}
			\label{tab:mq}
		\end{table}
				
		\subsubsection{Cavity Modes}
		
		Each mode in the cavity is designated by a couple $(p,q)\in\mathbb{N}^{*2}$, so that its radial wave number is $l_p$ and its vertical wave number is $m_q$. Consequently, a mode $(p,q)$ has a given frequency $\omega_{p,q}$ fixed by the linear dispersion relation for internal waves as
		\begin{equation}
			\frac{\omega_{p,q}}{N} = \left(\frac{l_p^2}{l_p^2 + m_q^2}\right)^{1/2}.\label{eq:IW38}
		\end{equation}
		 We present in table~\ref{tab:omegaspq} the values of $\omega_{p,q}/N$ corresponding to the lowest modes $(p,q)$ in a cavity of radius $R=20\mathrm{~cm}$ and height $L=60\mathrm{~cm}$.
		\begin{table}[!htbp]
			\centering
			\begin{tabular}{c||c c c c c c c c c}
		 		\hline 
		 		$\omega_{p,q}/N$ & $~~~q=1~~~$ & $~~~q=2~~~$ & $~~~q=3~~~$ & $~~~q=4~~~$ & $~~~q=5~~~$ & $~~~q=6~~~$ & $~~~q=7~~~$ & $~~~q=8~~~$ & $~~~q=9~~~$ \\ 
		 		\hline \hline
		 		$p=1$ & $0.9647$ & $0.8774$ & $0.7733$ & $0.6750$ & $0.5905$ & $0.5207$ & $0.4633$ & $0.4159$ & $0.3766$ \\ 
		 		$p=2$ & $0.9891$ & $0.9583$ & $0.9128$ & $0.8589$ & $0.8017$ & $0.7453$ & $0.6919$ & $0.6424$ & $0.5975$ \\
		 		$p=3$ & $0.9948$ & $0.9795$ & $0.9555$ & $0.9247$ & $0.8892$ & $0.8509$ & $0.8114$ & $0.7720$ & $0.7336$ \\
		 		$p=4$ & $0.9969$ & $0.9879$ & $0.9733$ & $0.9540$ & $0.9307$ & $0.9045$ & $0.8762$ & $0.8466$ & $0.8165$ \\
		 		$p=5$ & $0.9980$ & $0.9920$ & $0.9823$ & $0.9692$ & $0.9530$ & $0.9344$ & $0.9137$ & $0.8914$ & $0.8680$ \\
		 		\hline
		 		\end{tabular}  		
			\caption{Values of $\omega_{p,q}$ corresponding to modes $(p,q)$ of radial wave number $l_p$ and vertical wave number $m_q$ for $R=20\mathrm{~cm}$ and $L=60\mathrm{~cm}$.}
			\label{tab:omegaspq}
		\end{table}
		
		As shown in equation \eqref{eq:IW29}, a non-linear wave-wave interaction produces a wide range of waves that can be projected over the appropriate basis of linear solutions \eqref{eq:IW31}. With non-zero non-linear terms, as shown in equation \eqref{eq:IW26}, harmonic modes can be fed by an excitation mode. A mode at frequency $\omega$ is therefore in resonance with its $n$\textsuperscript{th} harmonic, $n\in\mathbb{N}^*$, if there exists a couple of non-zero integers $(p,q)$ such that
		\begin{equation}
			n \omega = \omega_{p,q}.\label{eq:IW39}
		\end{equation}
		In particular, a mode excited at frequency $\omega$ is in non-linear resonance with its first harmonic if twice the forcing frequency $2\omega$ is close to a frequency $\omega_{p,q}$ that corresponds to a mode $(p,q)$. We will verify that in the following sections.
		
	\section{Experimental Apparatus} 
	
\begin{figure}[!htbp]
			\centering
			\epsfig{file=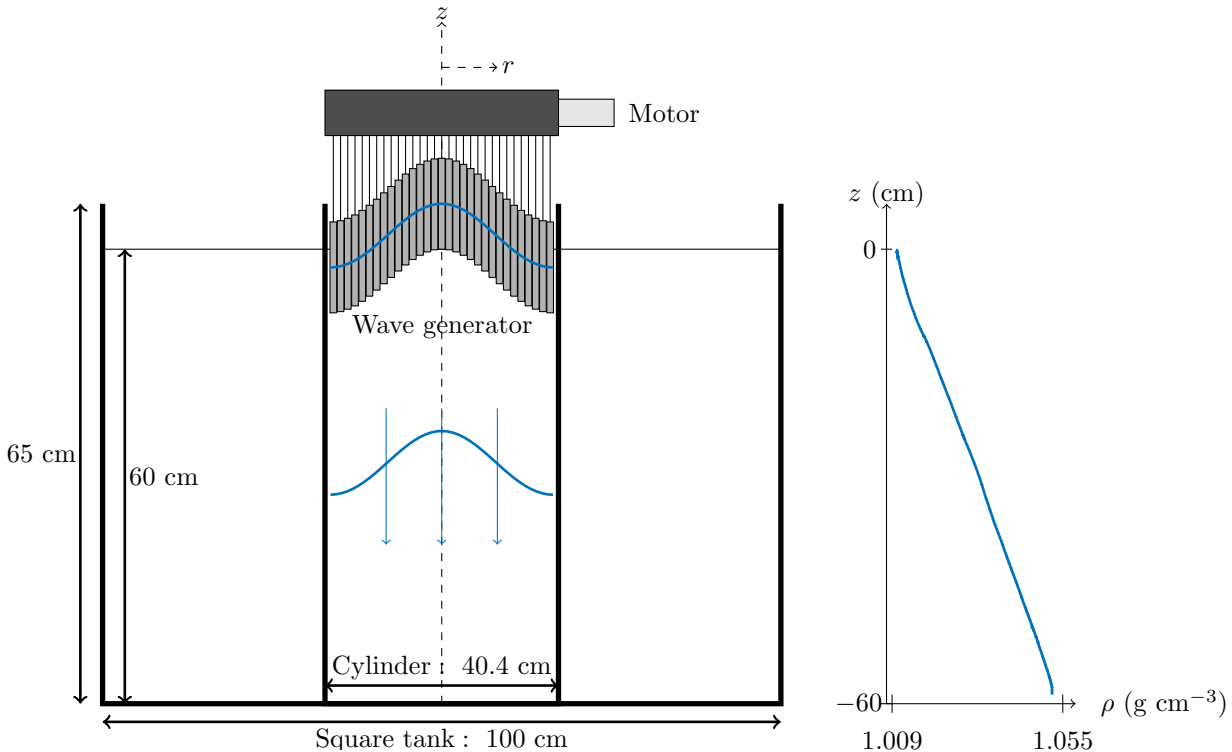}
			\caption{Schematic of the experimental apparatus. Left: a cylindrical tank, inside a square tank, confines the waves produced by the generator located at the surface, leading to a radial Bessel mode propagating downwards. Right: linear stratification measured in the experiments. Vertical dimension of the generator is not to scale.}
			\label{fig:fig1}
		\end{figure}

		Our experiments were conducted in the experimental apparatus described in Boury~\textit{et al.}~\cite{boury2018} and Boury~\textit{et al.}~\cite{boury2019}, adapted from the setup of Maurer \textit{et al}~\cite{maurer2017}. A general schematic of the experimental device is presented in figure~\ref{fig:fig1}. The system is described using natural cylindrical coordinates with the origin taken at the surface of the water at the center of the tank.
		
		The generator comprises sixteen, $12\mathrm{~mm}$ thick, concentric PVC cylinders periodically oscillating, each of them being forced by two eccentric cams. The eccentricities can be configured to introduce a phase shift between the different cylinders, and the oscillating amplitude can be set for each individual cylinder. As a result, the vertical displacement of the $n^{th}$ cylinder can be described by
		\begin{equation}
			a_n(t) = A_n \cos (\omega t + \alpha_n),\label{eq:IW40}
		\end{equation}
		with $A_n$ its amplitude, $\omega$ the forcing frequency, and $\alpha_n$ a phase shift. For a smooth motion of the PVC cylinders, a $1\mathrm{~mm}$ gap is kept between each cylinder and the total diameter of the wave generator is then $402\mathrm{~mm}$. The generator is mounted at the surface of the water to force downwards internal waves. The wave field is forced using a mode $1$ profile of radial wave number $l=19\mathrm{~m^{-1}}$, for which amplitudes of each cylinder are presented in table~\ref{tab:shafts}. This profile has been proven efficient to generate axisymmetric Bessel-shaped wave fields~\cite{boury2018}.
			\begin{table}[!htbp]
				\begin{center}
					\begin{tabular}{c c c c c c c c c c c c c c c c c c}
						\hline
						Cams & $1$ & $2$ & $3$ & $4$ & $5$ & $6$ & $7$ & $8$ & $9$ & $10$ & $11$ & $12$ & $13$ & $14$ & $15$ & $16$ \\
						\hline \hline
						Mode~$1$ amplitudes~$\mathrm{(mm)}$ & $2.5$ & $2.4$ & $2.3$ & $2.1$ & $1.9$ & $1.6$ & $1.3$ & $0.9$ & $0.6$ & $0.2$ & $-0.1$ & $-0.3$ & $-0.6$ & $-0.8$ & $-0.9$ & $-1$ \\
						\hline
					\end{tabular}
					\caption{Amplitudes of the different cams of the generator for a mode $1$ radial profile. The first cam is located at $r=0$.}
					\label{tab:shafts}
				\end{center}
			\end{table}
			
				Experiments were conducted in a transparent cylindrical acrylic tank of the same diameter as the generator, set into a square acrylic tank to prevent the experiment visualisation suffering from optical deformations that would occur due to curved interfaces. Both tanks were filled with salt-stratified water with the same density profile. We used the double-bucket method to fill the tanks with a linear stratification~\cite{fortuin1960, oster1963}. Density and buoyancy were measured as a function of depth using a calibrated PME conductivity and temperature probe mounted on a motorised vertical axis. Buoyancy frequency was estimated from the mean value of the $N$ profile obtained from the density function $\rho(z)$ measured from the free surface to within a couple of centimeters of the bottom of the tank, due to the construction of the probe. The wave generator was immersed at a depth of $2\mathrm{~cm}$ into the stratification. The error on the buoyancy frequency was estimated using the standard deviation of this $N$ profile, which was about $4\%$ of the estimated $N$ value (see~\cite{boury2018} for more details). We obtained a buoyancy frequency of $N = 0.89\pm 0.06\mathrm{~rad\cdot s^{-1}}$.
		
		Velocity fields were obtained via Particle Image Velocimetry (PIV). A laser sheet was created by a laser beam (Ti:Sapphire, $2\mathrm{~watts}$, wavelength $532\mathrm{~nm}$) going through a cylindrical lens. It could be oriented either horizontally (to measure the radial and orthoradial velocity) or vertically (to measure the vertical and the radial velocity). For the purpose of visualisation, $10\mathrm{~\mu m}$ diameter hollow glass spheres of volumetric mass $1.1\mathrm{~kg\cdot L^{-1}}$ were added to the fluid while filling the tank. To obtain good quality velocity fields near the bottom of the tank and while imaging in a horizontal plane, $10\mathrm{~\mu m}$ silver-covered spheres of volumetric mass $1.4\mathrm{~kg\cdot L^{-1}}$ were added when needed in some experiments. Images were recorded at $4\mathrm{~Hz}$ and data processing of the PIV raw images was done using the CIVx algorithm~\cite{fincham2000}.
		
	\section{Results} 
	
		\subsection{Axisymmetric Non-Linear Wave Generation}
		
\begin{figure}[!htbp]
			\centering
			\epsfig{file=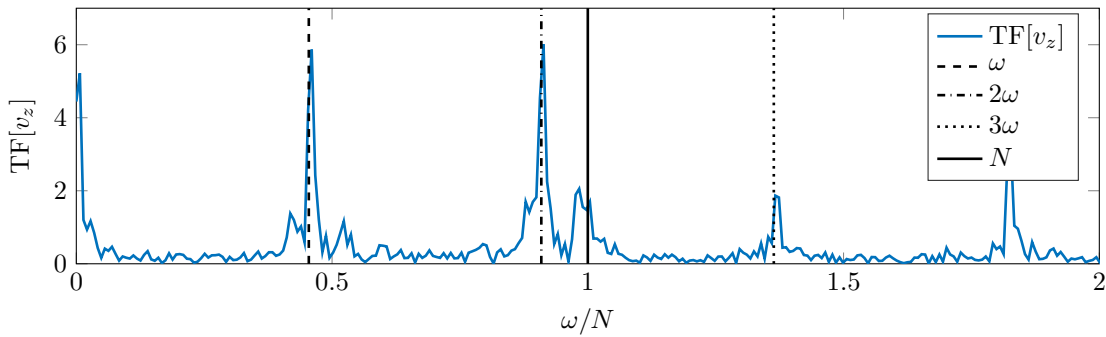}
			\caption{Example of Fourier transform performed over the last two minutes of experiment $9$ with a forcing at $\omega/N=0.449$. The solid line shows the cut-off frequency at $N$. Dashed, dashed-dotted, and dotted lines show $\omega$, $2\omega$, and $3\omega$ frequencies.}
			\label{fig:fig2}
		\end{figure}

		
			We consider a set of nine experiments indexed from $0$ to $9$, with forcing frequency from $\omega/N = 0.305$ to $0.449$, using a mode $1$ configuration at the generator~\cite{boury2018}. For each frequency, a $10$ minute forcing leads to non-linear wave-wave interactions, where higher frequency waves are created. This phenomenon is illustrated in figure~\ref{fig:fig2}, showing the frequency spectrum from experiment number $9$, forced at $\omega/N=0.449$. The spectrum contains not only the forcing frequency at $\omega/N$, but also a mean flow at $\omega=0$ and several harmonics at $2\omega/N$, $3\omega/N$, and so on. Here, the first harmonic would be propagating as $\omega/N < 0.5$ and therefore $2\omega/N < 1$. This behaviour is observed for all frequencies, regardless of any resonant cavity aspect identified in Boury~\textit{et al.}~\cite{boury2018}. In the performed experiments, due to the range of frequencies choosen, the first harmonic at $2\omega/N$ is always propagating and, except for experiments $1$ and $2$, the second harmonic at $3\omega/N$ is evanescent.
	
\begin{figure}[!htbp]
		\centering
		\epsfig{file=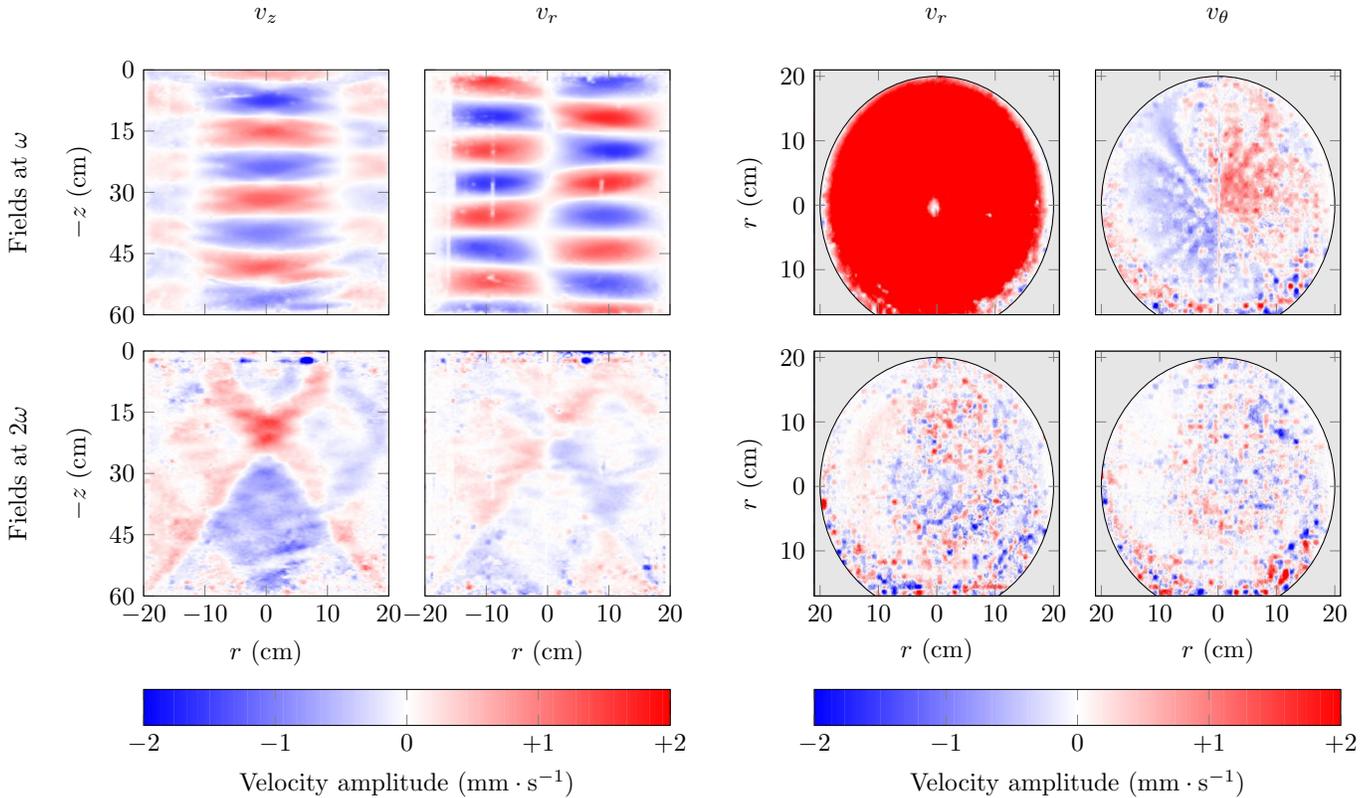, width=\textwidth}
		\caption{Velocity fields in experiment $9$, at forcing frequency $\omega/N = 0.449$. The first row shows the filtered wave field at $\omega$ and the second row shows the filtered wave field at $2\omega$. From left to right: $v_z$, $v_r$ in a vertical cross-section, and $v_r$, $v_\theta$ in a horizontal cross section. For the purpose of visualisation, negative values of $r$ are used in the vertical plane views.}
		\label{fig:fig3}
	\end{figure}

			Figure~\ref{fig:fig3} gives snapshot examples of such an experiment, for a forcing at $\omega/N=0.449$, corresponding to the spectrum presented in figure~\ref{fig:fig2}. The first row shows the wave field filtered at the forcing frequency $\omega$ with, from left to right, $v_z$ and $v_r$ in the vertical plane, and $v_r$ and $v_\theta$ in the horizontal plane. The excited wave field corresponds to a mode $1$ and presents all the features of an axisymmetric wave field~\cite{boury2018}: right-left symmetry of $v_z$ and right-left antisymmetry of $v_r$ in the vertical plane, invariance by rotation of center $(0,0)$ for $v_r$ in the horizontal plane, and a negligible orthoradial velocity $v_\theta$ when observed in the horizontal plane ($v_r$ dominates $v_\theta$ by a factor $\sim 10$ in magnitude).
			
			The first harmonic created at $2\omega/N$ is filtered and shown in the second row of figure~\ref{fig:fig3}. Though no spatial wave lengths can be directly infered from the wave field, we notice that the created harmonic field shares the same axisymmetric properties as the excitation wave field. The right-left symmetry/antisymmetry of $v_z$ and $v_r$ in the vertical plane is also consistent with an axisymmetric description of the wave field. More specifically, a zero of radial velocity is observed at the center of the experimental domain whereas filtered wave fields in Triadic Resonant Instability (TRI)~\cite{maurerPhD} have shown, in some cases, non-zero velocity, breaking the axisymmetry of the system. The orthoradial velocity $v_\theta$ is, as compared to the excitation wave field, a noisy signal and does not show any sign of symmetry breaking.
			
		\subsection{Proper Orthogonal Decomposition (POD)}
		
			In order to describe quantitatively the wave field at a given frequency and, more particularly, the wave field created in the harmonic at twice the forcing frequency, we used a Proper Orthogonal Decomposition (POD) method. As we know the relevant stream functions (modes) in the cavity, the POD process simply consists of projecting the experimental wave field over an orthogonal basis built accordingly.
			
			The series of $\left\lbrace r \mapsto J_0(l_p r)~|~p \in \mathbb{N}^* \right\rbrace$ and $\left\lbrace z \mapsto \sin(m_q z)~|~q \in \mathbb{N}^* \right\rbrace$ forms an orthogonal basis in an axisymmetric geometry, which is appropriate for this study. A direct consequence is that the stream function can be written in terms of normalised modes $(p,q)$, denoted $\psi_{p,q}$, in the axisymmetric domain $(r,\theta,z) \in \mathcal{C} = [0;~R]\times[0;~2\pi]\times[0;~-L]$, defined as
			\begin{equation}
				\psi_{p,q} (r,z) = \psi_{p,q}^0 J_1 (l_p r) \sin(m_q z),\mathrm{~~~~~~~with~~~~~~~}\psi_{p,q}^0 = \frac{1}{J_1' (l_p R)} \sqrt{\frac{2}{L \pi}},\label{eq:IW41}
			\end{equation}
			satsifying the condition
			\begin{equation}
				\int_\mathcal{C} \left[\psi_{p,q} (r,z)\right]^2 r \diff r \,\diff\theta \,\diff z = 1,\label{eq:IW42}
			\end{equation}
			from which we can derive the vertical and radial velocities associated to a mode $(p,q)$, using equations \eqref{eq:IW5}, as follows
			\begin{eqnarray}
				v_{z}^{p,q} (r,z) &=& \frac{1}{r}\dfrac{\partial (r \psi_{p,q})}{\partial r} = - l_p \psi_{p,q}^0 J_0 (l_p r) \sin(m_q z),\label{eq:IW43}\\
				v_{r}^{p,q} (r,z) &=& -\frac{1}{r}\dfrac{\partial (r \psi_{p,q})}{\partial z} = m_q \psi_{p,q}^0 J_1 (l_p r) \cos(m_q z).\label{eq:IW44}
			\end{eqnarray}
			Hence, the kinetic energy contained in one mode $(p,q)$  is given by
			\begin{equation}
				K_{p,q} = \left[ \int_\mathcal{C} v_z(r,z) \cdot v_{z}^{p,q}(r,z) r \diff r \,\diff \theta \,\diff z \right]^2 + \left[ \int_\mathcal{C} v_r(r,z) \cdot v_{r}^{p,q}(r,z) r \diff r \diff \,\theta \,\diff z \right]^2,\label{eq:IW45}
			\end{equation}
			which, accounting for the discretisation of our domain, can be written as follows
			\begin{equation}
				K_{p,q} =  \left[ \sum_{r,z} 2\pi r v_z(r,z) \cdot v_{z}^{p,q}(r,z)\right]^2  + \left[ \sum_{r,z} 2 \pi r v_r(r,z) \cdot v_{r}^{p,q}(r,z)\right]^2,\label{eq:IW46}
			\end{equation}
			with sums over all spatial grid points. As a result, if we denote $K_0$ the kinetic energy of the total wave field, the fraction of energy in a mode $(p,q)$ is given by the scalar product \eqref{eq:IW46} normalised by $K_0$. The higher the quantity, the more dominant the mode is in the observed field. Note that the prefactors for $v_{z}^{p,q}$ and $v_{r}^{p,q}$ can differ if the normalisation process is directly applied to the stream function or to the velocities, though it will not affect the conclusions.
			
\begin{figure}[!htbp]
				\centering
				\epsfig{file=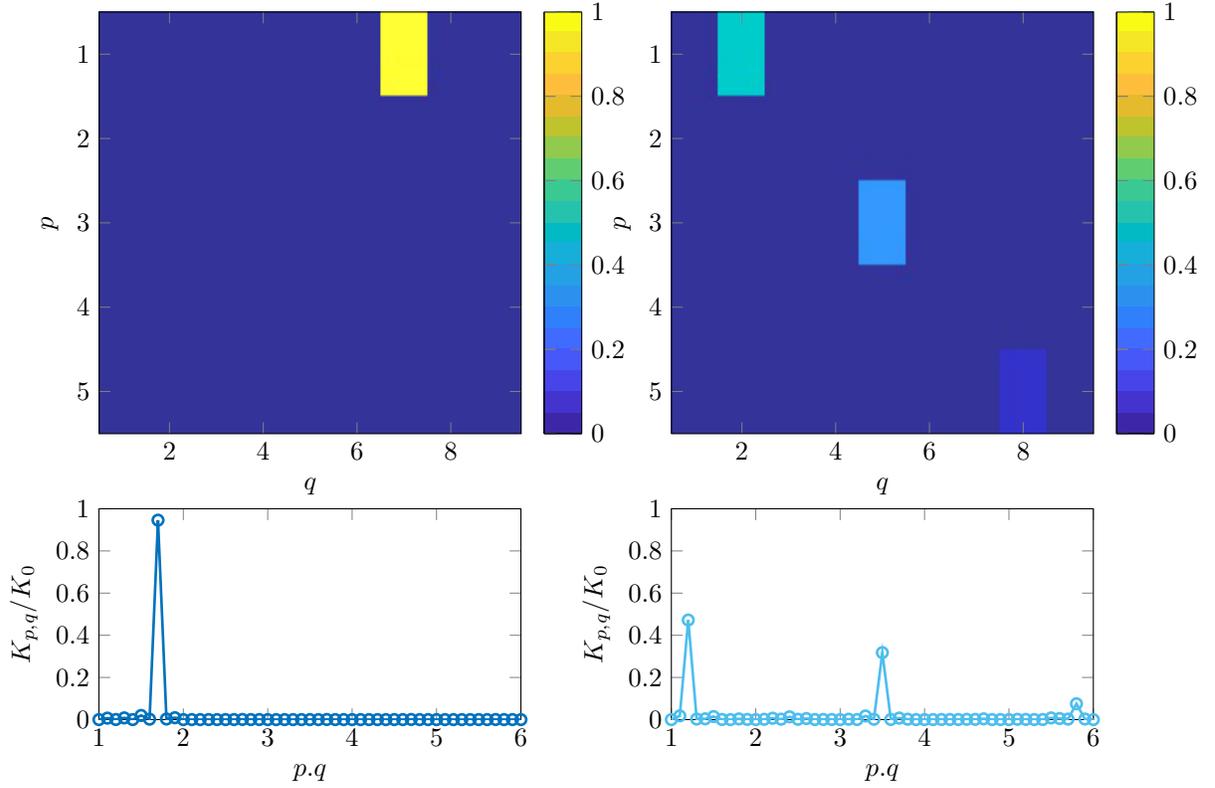}
				\caption{POD performed on experiment number $9$. The top row is the colormap of the energy distribution in the different modes in the $(p,q)$ plane, and the bottom row is the transposition of this $2$D plot into a $1$D plot of axis $p.q=p+0.1 q$. Left: POD over the filtered wave field at the excitation frequency $\omega/N = 0.449$. Right: POD over the filtered wave field at the first harmonic frequency $2\omega/N = 0.898$.}
				\label{fig:fig4}
			\end{figure}

			The left panel of figure~\ref{fig:fig4} shows an example of such a POD decomposition using the excitation wave field previously presented in the top part of figure~\ref{fig:fig3}, and the right panel of figure~\ref{fig:fig4} shows the decomposition of the first harmonic wave field, respectively at $\omega/N = 0.449$ and $2\omega/N = 0.898$. Figure~\ref{fig:fig4} (top) depicts the colormap of the energetic distribution onto different modes and figure~\ref{fig:fig4} (bottom) plots the kinetic energy contained in every tested mode as a function of the quantity $p.q$, defined as $p.q=p+0.1 q$, used to identify the different modes with a single number or, in other words, to transform the $2$D plot of the top row of figure~\ref{fig:fig4} into the more easily quantifiable $1$D plot of the bottom of figure~\ref{fig:fig4}. As expected, figure~\ref{fig:fig4} (left) shows that nearly all the energy of the wave field produced by the generator lies in a radial mode $1$ wave, more exactly a $(1,7)$ mode, that can be clearly identified in figure~\ref{fig:fig3} with $7$ zeros along the vertical direction and one along the horizontal direction. In addition, while the filtered wave field at $2\omega/N$ in figure~\ref{fig:fig3} does not explicitly display a mode, we see from figure~\ref{fig:fig4} (right) that the energy is mostly split into mode $(1,2)$ ($47\%$), mode $(3,5)$ ($32\%$), and mode $(5,8)$ ($8\%$), the other contributions being negligible. The first harmonic generated in experiment $9$ can therefore be described as a sum of modes $(1,2)$, $(3,5)$, and $(5,8)$, with given prefactors to account for the distribution of energy between the three modes. 
		
		\subsection{Harmonics and Mode Selection}
		
\begin{figure}[!htbp]
				\centering
				\epsfig{file=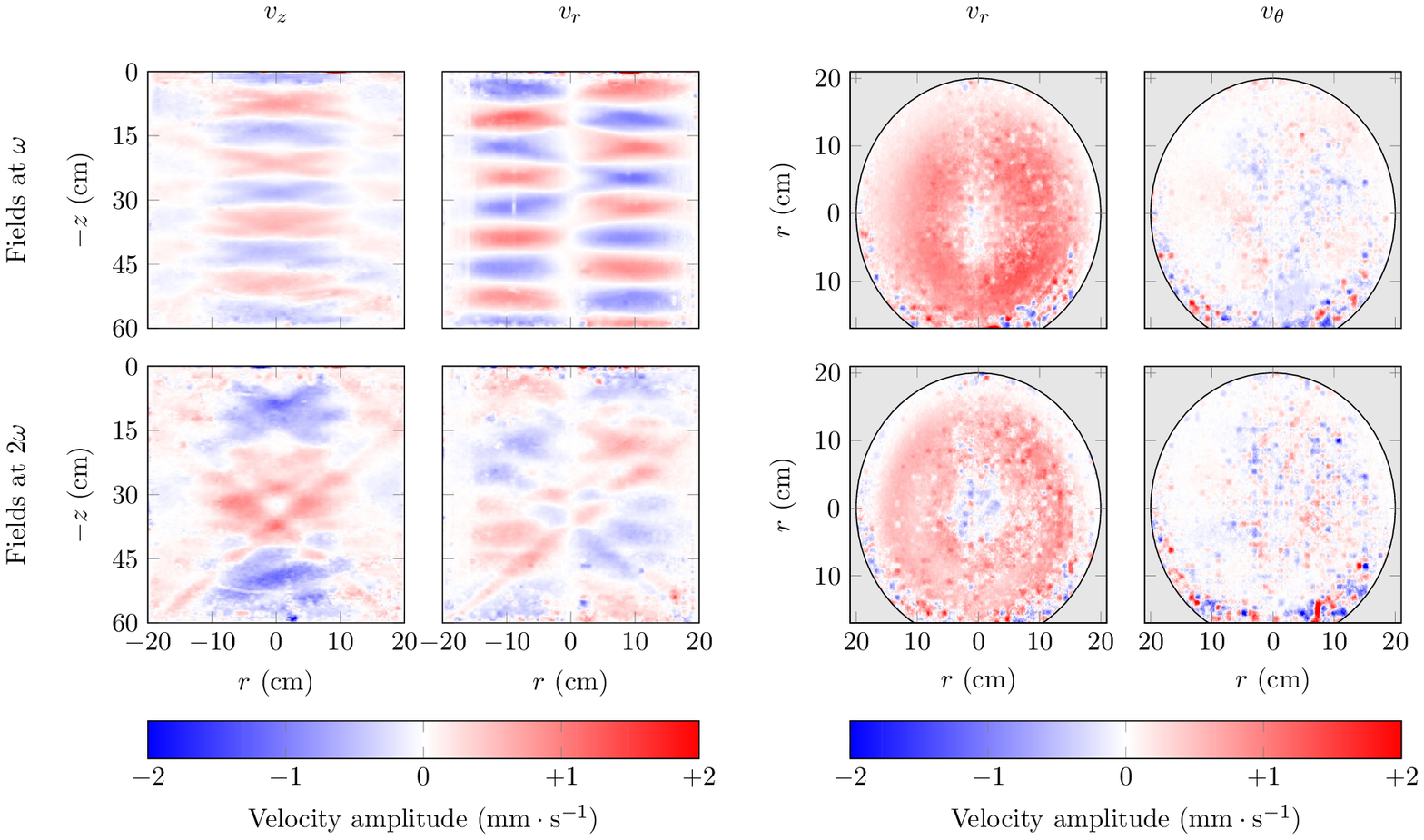, width=\textwidth}
				\caption{Velocity fields in experiment $6$, with $2\omega/N = 0.800$, with generation of a mode $(1,3)$.}
				\label{fig:fig5}
			\end{figure}

\begin{figure}[!htbp]
				\centering
				\epsfig{file=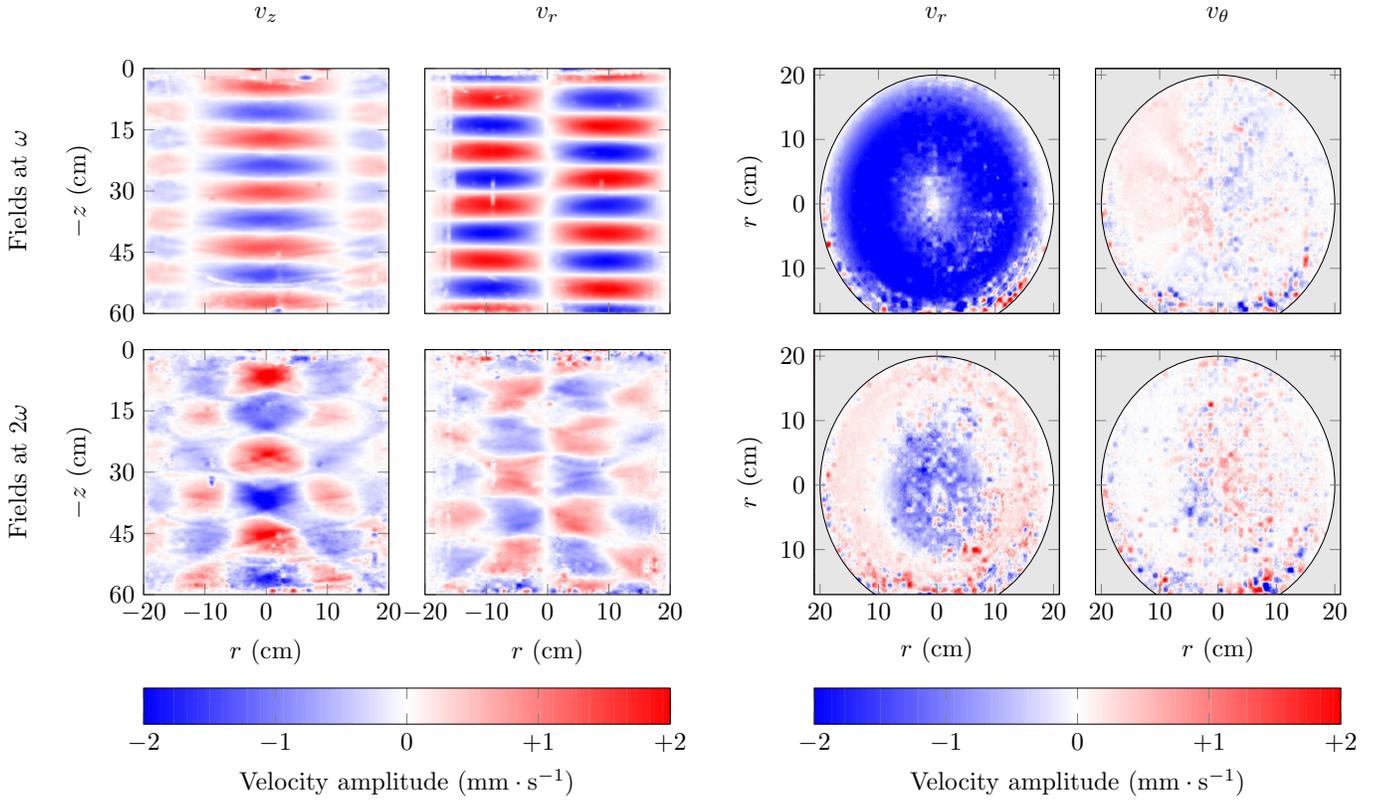, width=\textwidth}
				\caption{Velocity fields in experiment $5$, with $2\omega/N = 0.765$, with generation of a mode $(2,6)$.}
				\label{fig:fig6}
			\end{figure}

\begin{figure}
				\centering
				\epsfig{file=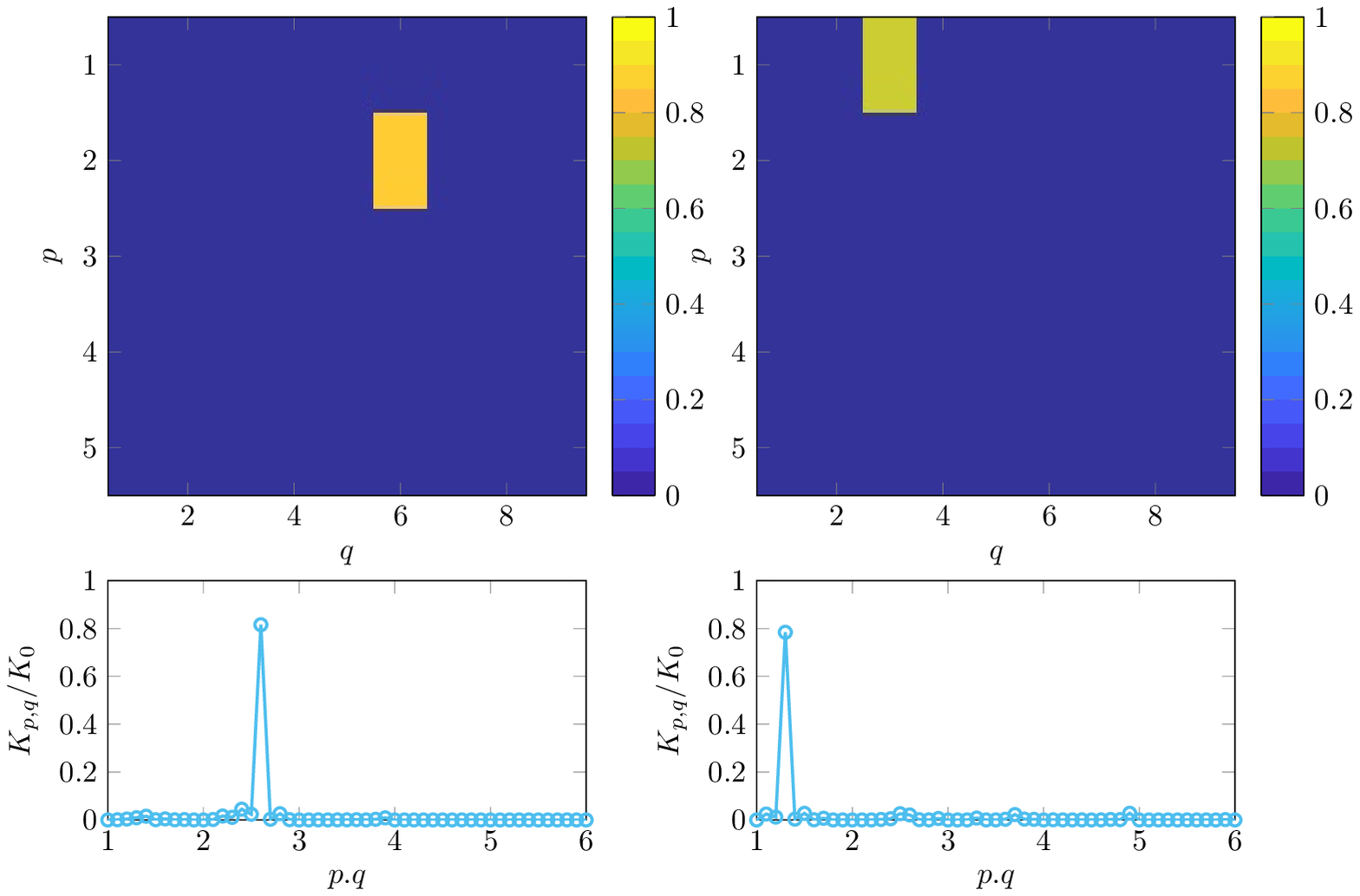}
				\caption{POD projection for experiments $5$ (left) and $6$ (right). The top row is the colormap of the energy distribution in the different modes in the $(p,q)$ plane, and the bottom row is the transposition of this $2$D plot into a $1$D plot of axis $p.q=p+0.1 q$.}
				\label{fig:fig7}
			\end{figure}

			The POD discrimination process is applied to experiments $1$ to $9$ and its results are summarised in table~\ref{tab:harmonics}. In some cases, the first harmonic generated by the non-linear wave-wave interaction can be clearly identified by eye as a single mode $(p,q)$, as shown for experiments $5$ (with $2\omega/N=0.755$, see figure~\ref{fig:fig5}) and $6$ (with $2\omega/N=0.791$, see figure~\ref{fig:fig6}), with the generation of a mode $(1,3)$ and a mode $(2,6)$, respectively. In such cases, this clearly observed mode is the one identified as the most energetic in the POD decomposition. Figure~\ref{fig:fig7} illustrates this point with experiment $5$ and $6$, showing a single mode containing about $80$ to $90\%$ of the total kinetic energy of the first harmonic. The remaining fraction of energy, from $10$ to $20\%$, is sparsely distributed into lower modes that contain less than $5\%$ of the total energy each and do not contribute significantly to the general form and behaviour of the wave field. In other cases, as described in the previous section for experiment $9$, the energy is more evenly distributed between several modes.
	
			The structure of the generated harmonic modes does not always match the one of the excitation mode. For example, experiments $3$ and $6$ lead to generation of a dominant mode with a radial structure similar to the excitation wave field ($l_1=19\mathrm{~m^{-1}}$), whereas experiments $2$, $5$, and $8$, generate a dominant higher order radial structure ($l_2 = 35\mathrm{~m^{-1}}$). For a given radial structure, we note that the vertical wave number increases with the frequency $2\omega/N$, consistently with the dispersion relation \eqref{eq:IW19}. Interestingly, the mode frequencies are not ordered by lower $p$ or lower $q$ but seem randomly distributed: for example, by increasing $\omega$, the harmonics can be dominated by a radial mode $1$, then a radial mode $2$, and then a radial mode $1$ again.
		
		\begin{table}[!htbp]
		\centering
		\begin{tabular}{c c c l l l}
			\hline 
			Experiment & $~~~\omega/N~~~$ & $~~~2\omega / N~~~$ & $(p,q)~[\%]~$ & & \\ 
			\hline \hline
			$1$ & $0.305$ & $0.611$ & $(1,5)~[61\%]~$ & $(2,9)~[18\%]~$ & \\
			$2$ & $0.323$ & $0.647$ & $(2,8)~[61\%]~$ & $(1,4)~[10\%]~$ & \\
			$3$ & $0.341$ & $0.683$ & $(1,4)~[89\%]~$ & & \\
			$4$ & $0.359$ & $0.719$ & $(1,4)~[22\%]~$ & $(2,7)~[21\%]~$ & \\
			$5$ & $0.377$ & $0.755$ & $(2,6)~[82\%]~$ & & \\
			$6$ & $0.395$ & $0.791$ & $(1,3)~[78\%]~$ & & \\
			$7$ & $0.413$ & $0.827$ & $(2,5)~[35\%]~$ & $(1,3)~[14\%]~$ & $(3,7)~[8\%]~$ \\
			$8$ & $0.431$ & $0.863$ & $(2,4)~[69\%]~$ & $(3,6)~[9\%]~$ & \\
			$9$ & $0.449$ & $0.899$ & $(1,2)~[47\%]~$ & $(3,5)~[32\%]~$ & $(5,8)~[8\%]~$ \\
			\hline 
		\end{tabular} 		
		\caption{Forcing frequencies and their first harmonics, with identified dominating modes (in $\%$ of kinetic energy) using POD.}
		\label{tab:harmonics}
		\end{table}
		
		\begin{table}[!htbp]
			\centering
			\begin{tabular}{c||c c c c c c c c c}
		 		\hline 
		 		$\omega_{p,q}/N$ & $~~~q=1~~~$ & $~~~q=2~~~$ & $~~~q=3~~~$ & $~~~q=4~~~$ & $~~~q=5~~~$ & $~~~q=6~~~$ & $~~~q=7~~~$ & $~~~q=8~~~$ & $~~~q=9~~~$ \\ 
		 		\hline \hline
		 		$p=1$ & $0.9647$ & \textcolor{mycolor3}{$\mathbf{0.8774}$ [9]} & \textcolor{mycolor3}{$\mathbf{0.7733}$ [6,7]} & \textcolor{mycolor3}{$\mathbf{0.6750}$ [2,3,4]} & \textcolor{mycolor3}{$\mathbf{0.5905}$ [1]} & $0.5207$ & $0.4633$ & $0.4159$ & $0.3766$ \\ 
		 		$p=2$ & $0.9891$ & $0.9583$ & $0.9128$ & \textcolor{mycolor3}{$\mathbf{0.8589}$ [8]} & \textcolor{mycolor3}{$\mathbf{0.8017}$ [7]} & \textcolor{mycolor3}{$\mathbf{0.7453}$ [5]} & \textcolor{mycolor3}{$\mathbf{0.6919}$ [4]} & \textcolor{mycolor3}{$\mathbf{0.6424}$ [2]} & \textcolor{mycolor3}{$\mathbf{0.5975}$ [1]} \\
		 		$p=3$ & $0.9948$ & $0.9795$ & $0.9555$ & $0.9247$ & \textcolor{mycolor3}{$\mathbf{0.8892}$ [9]} & \textcolor{mycolor3}{$\mathbf{0.8509}$ [8]} & \textcolor{mycolor3}{$\mathbf{0.8114}$ [7]} & $0.7720$ & $0.7336$ \\
		 		$p=4$ & $0.9969$ & $0.9879$ & $0.9733$ & $0.9540$ & $0.9307$ & $0.9045$ & $0.8762$ & $0.8466$ & $0.8165$ \\
		 		$p=5$ & $0.9980$ & $0.9920$ & $0.9823$ & $0.9692$ & $0.9530$ & $0.9344$ & $0.9137$ & \textcolor{mycolor3}{$\mathbf{0.8914}$ [9]} & $0.8680$ \\
		 		\hline
		 		\end{tabular}  		
			\caption{Reproduction of table~\ref{tab:omegaspq}, showing the modes $(p,q)$ identified in the POD. Bold red values of $\omega_{p,q}/N$ are the theoretical frequencies corresponding to the modes observed in the experiments. The associated experiments are indicated between brackets.}
			\label{tab:tab6}
		\end{table}
		
		Table~\ref{tab:tab6} reproduces the theoretical results of table~\ref{tab:omegaspq}, highlighting the frequencies corresponding to the observed modes of experiments $1$ to $9$, which can be compared to the values in table~\ref{tab:harmonics}. This comparison shows that, when a dominant mode $(p,q)$ is observed, the frequency of the harmonic (third column in table~\ref{tab:harmonics}) is close to the frequency of the same mode $(p,q)$ stated by the dispersion relation (table~\ref{tab:tab6}). For instance, in experiment $6$, $78\%$ of the energy of the super-harmonic is in a mode $(1,3)$, which is the mode observed in figure~\ref{fig:fig5}, and the frequency predicted for such a mode is $\omega_{1,3}/N = 0.773$ according to table~\ref{tab:tab6}, close to the harmonic frequency observed in the experiment $2\omega/N=0.791$ (table~\ref{tab:harmonics}). Similarly, in experiment $5$, $82\%$ of the energy of the super-harmonic is in a mode $(2,6)$, which is the mode observed in figure~\ref{fig:fig6}, and the frequency predicted for such a mode is $\omega_{2,6}/N = 0.745$, also close to the harmonic frequency observed in the experiment $2\omega/N=0.755$.
		
\begin{figure}[!htbp]
				\centering
				\epsfig{file=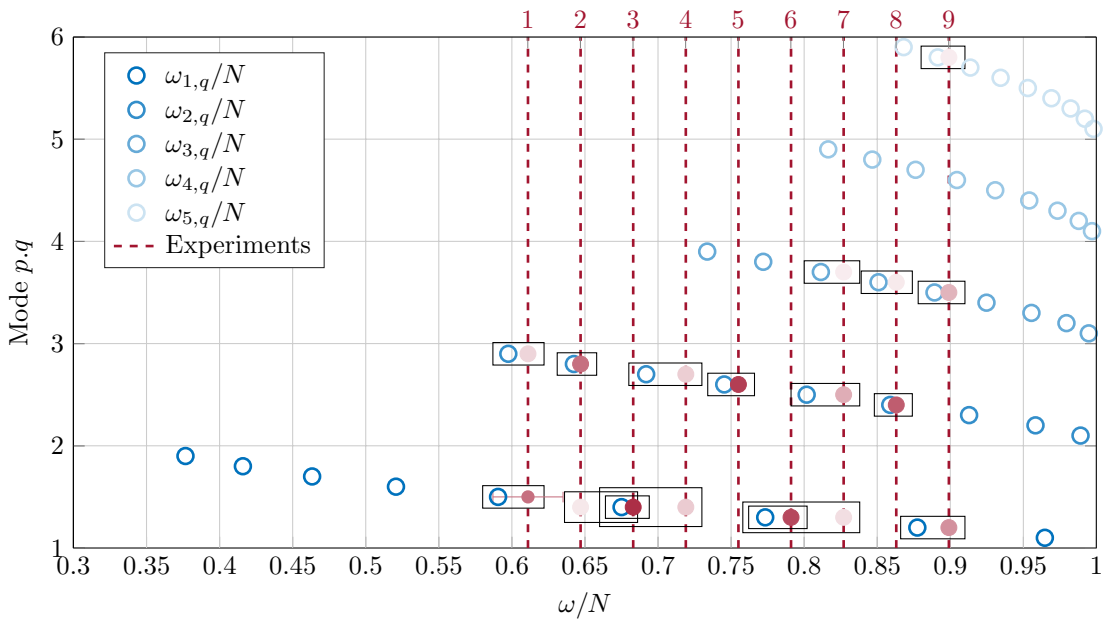}
				\caption{Selected modes vs. frequency. The blue circles show the frequencies and mode numbers of the cavity modes $(p,q)$ (from table~\ref{tab:omegaspq}). The red dashed lines show the first harmonics $(2\omega/N)$ seen in the $9$ conducted experiments, which numbers are indicated above each line. On each line, red dots indicate the modes identified with POD, darker red meaning a more energetic mode. For each experimentally observed mode (dots), the corresponding theoretical mode (circles) is indicated by a linking rectangle. The error on $N$ is illustrated by an error bar in experiment $1$.}
				\label{fig:fig8}
			\end{figure}

		To generalize this observation, we summarised our results in figure~\ref{fig:fig8}. The blue circles show, in the phase space $(x=\omega/N, y=p.q)$, all the theoretical cavity modes that can be created according to table~\ref{tab:omegaspq} (for $p=1$ through $5$ and $q=1$ through $9$). In parallel, each first harmonic frequency for experiment $1$ to $9$ is represented by a vertical dashed line at constant $\omega/N$. At each of these frequencies, modes extracted from the POD projection of the filtered harmonic wave fields are represented by red dots in the phase space, where the color intensity shows the energetic contribution of each dominant mode. To guide the eye, black rectangles link together an experimentally observed mode with its theoretical counterpart whose frequency is within $1$ to $7\%$ of the experimental frequency in all cases. This margin is of the order of the error on the buoyancy frequency $N$ measured with the probe (about $4\%$). On a vertical dashed line, when a single red dot is plotted (see for example experiment 5), it means that the harmonic wave field looks like a single mode which, as shown by the linking rectangle, corresponds to the mode selected by the non-linear interaction. On the contrary, when several dots appear for a given harmonic frequency (see, for example, experiment 1), resonant conditions stated by the dispersion relation are not exactly fulfilled and the harmonic wave field is a combination of two or more modes, the system being unable to select a solitary one.
		
	\section{Conclusions and Discussion}
	
		This study presents, to our knowledge, the first experimental evidence of internal wave harmonics created by weakly non-linear internal wave self-interaction in a linear stratification. We have presented a configuration in which, as the excited wave field has a frequency $\omega$ below $N/2$, the first harmonic created can always propagate and we have observed that the experimental PIV velocity fields display coherent structures when filtered at the harmonic frequency $2\omega/N$. With a simple analytical argument, we have shown that the confined domain provides restrictive boundary conditions so that the non-linear wave field has to satisfy a given set of constraints, leading to a discrete collection of radial and vertical wave numbers that can be theoretically predicted. The modes, or standing waves, defined by these relations constitute an excellent description of the first harmonic field in terms of a sum of several contributions, sometimes with a clear dominant one.
		
		Interestingly, self-interacting non-linear terms (the Jacobians) are null when computed using plane waves in Cartesian geometry. Hence, such a non-linear interaction cannot exist in linearly stratified fluids and, to obtain super-harmonics as in~\cite{baker2020}, a non-linear buoyancy frequency profile is required so that non-linearities can excite waves at higher frequencies through a second order forcing term. This study shows that for pure axisymmetric wave fields, however, self-interaction and super-harmonic generation is a common process even within linear stratifications.
		
		Through a Proper Orthogonal Decomposition (POD) technique, we have shown that a mode selection occurs while generating the harmonic wave field. For some frequencies, the first harmonic reproduces almost perfectly a mode, both in the vertical and in the radial directions and, in other cases, the first harmonic is a sum of modal wave fields. Moreover, in all cases, the newly generated wave field remains axisysmmetric. Modes of lowest radial wave numbers seem likely to be selected through this weakly non-linear interaction and, indeed, the most energetic modes are found to be for $p=1$ and $p=2$, whereas the few contributions observed at $p=3$ and beyond are usually small. This behaviour can be explained by the resistance of the stratification against vertical motions, which increases the efficiency of energy transfers to lower radial modes and, especially, to a mode $p=1$ whose structure is already forced by the wave generator. Note that higher modes, for $p>5$ and $q>9$, could also exist in the harmonic wave field but, for clarity, they are not displayed in figure~\ref{fig:fig8}. When tested using the POD decomposition, all of these modes showed very little contribution to the overall kinetic energy (about a few percent maximum).
		
		From a theoretical point of view, the exact generation process is not fully understood yet. A more detailed study, using Green functions could be undertaken to derive the exact forcing terms that generate super-harmonics, but falls beyond the scope of this work. Moreover, in such an experiment, other non-linear phenomena such as wave breaking or Triadic Resonant Instability (TRI), might occur. The underlying mechanisms, however, are likely to be different as, for example, they yield to the generation of sub-harmonics in the case of TRI, or they may involve a symmetry breaking~\cite{maurerPhD} and a fully developped cylindrical wave field with potential orthoradial velocities, which are still to be explained.

\medskip
\textbf{Acknowledgements:}
	
	This work has been partially supported by the ANR through grant ANR-17-CE30-0003 (DisET) and by ONR Physical Oceanography Grant N000141612450. S.B. wants to thank Labex iMust for supporting his research. This work has been achieved thanks to the resources of PSMN from ENS de Lyon.

\appendix
\section{Axisymmetric Green Function and Weakly Non-Linear Asymptotics}
		
			Inspired by the approach of \cite{voisin2003, ermanyuk2011, voisin2011}, we compute the Green function of the linear operator $\mathcal{L}$ in cylindrical coordinates (for more details on the calculus, see \cite{alastuey2008, nist2010}). In the axisymmetric case, without $\theta$-dependence, the Green function of the linear wave operator \eqref{eq:IW15} can be written as follows
			\begin{equation}
				G(\mathbf{r},t; \mathbf{r'},t') = \frac{G_0}{4\pi^2}\int _{-\infty}^{+\infty} e^{i \omega (t-t')} \int _{-\infty}^{+\infty} e^{i m (z-z')} [J_1(l r_{<}) Y_1(l r_{>})] \diff m \diff \omega,
			\end{equation}
			where $G_0$ is a normalisation coefficient, and $\omega$, $l$, and $m$ satisfy the dispersion relation \eqref{eq:IW19}. The radial dependence is expressed through the first order Bessel functions $J_1$ and $Y_1$, two analytical solutions of the radial part of the wave equation \cite{boury2018}. The variables are noted $t$, $\mathbf{r} = (r,\theta,z)$, $t'$, and $\mathbf{r'} = (r',\theta',z')$, and we use the notation $r_{<} = \min(r,r')$ and $r_{>} = \max(r,r')$.

			From the wave equation and the top boundary forcing, the linear wave field can be obtained as a solution of the problem
			\begin{equation}
				\left\{
				    \begin{array}{ll}
				        \mathcal{L}[\psi,b] = 0 & \mathrm{~for~} \mathbf{r} \in \mathcal{C} \\
				        \psi = \delta(z) J_0(l r)\cos(\omega t) & \mathrm{~for~} \mathbf{r} \in \partial\mathcal{C},
    					\end{array}
				\right.\label{eq:appendix2}
			\end{equation}
			where $\mathcal{C} = \left\lbrace(r,\theta,z) \in [0;~R]\times[0;~2\pi]\times[0;~-L]\right\rbrace$ is the cavity domain and $\partial\mathcal{C}$ stands for its boundaries. The forcing at the top boundary is expressed by a Dirac distribution $\delta(z)$. Equation ~\ref{eq:appendix2} admits for solution
			\begin{equation}
				\psi (z,r,t) = \int_{\partial\mathcal{C}\times\mathbb{R}} \delta(z') J_0 (l r') \cos (\omega t') \dfrac{\partial G}{\partial \mathbf{n}}(\mathbf{r},t; \mathbf{r'},t') \diff S' \diff t',
			\end{equation}
			while integrating over the boundary $\partial\mathcal{C}$.
			
			A small parameter $\varepsilon=\psi_0 T /L^2$ characterising the contribution of non-linear terms can be defined from the derivation proposed in \citep{boury2019} with $\psi_0$ the stream function amplitude, $T$ and $L$ characteristic temporal and spatial periods, allowing to write the stream function with a similar development as in Husseini~\textit{et al.}~\cite{husseini2019} in the following expansion
			\begin{equation}
				\psi = \varepsilon \psi^{(1)} + \varepsilon^2 \psi^{(2)} + \mathcal{O}\left(\varepsilon^3\right),
			\end{equation}
			where $\psi^{(1)}$ is the solution to the linear problem previously discussed. Considering that the second order terms come from the non-linear term
			\begin{equation}
				\mathcal{N} [\psi^{(1)}, b^{(1)}] = \varepsilon^2 C J_1 (l r) \left[ J_2 (l r) \sin^2 (\omega t - m z) - J_1 (l r) \cos^2 (\omega t - m z) \right],
			\end{equation}
			at lowest order, the weakly non-linear correction $\psi^{(2)}$ can be found by defining a new Green problem where the non-linear self-interaction of the linear term acts as a forcing field over the whole domain
			\begin{equation}
				\left\{
				    \begin{array}{ll}
				        \mathcal{L}[\psi^{(2)},b^{(2)}] = \varepsilon^2 \mathcal{N}[\psi^{(1)}, b^{(1)}] & \mathrm{~for~} \mathbf{r} \in \mathcal{C} \\
				        \psi = 0 & \mathrm{~for~} \mathbf{r} \in \partial\mathcal{C},
    					\end{array}
				\right.
			\end{equation}
			whose solution is found analytically as
			\begin{equation}
				\varepsilon^2 \psi^{(2)} (r,z,t) = \varepsilon^2 \int_{\mathcal{C}\times\mathbb{R}} \mathcal{N}[\psi^{(1)}, b^{(1)}] G(\mathbf{r},t; \mathbf{r'},t') \diff r' \diff z' \diff t'.
			\end{equation}			
			This expression predicts that the wave created through self-interaction is a first harmonic at $2\omega$, or a mean flow at zero frequency. The spatial dependence, however, being coupled between $l$ and $m$, cannot be directly infered.

\bibliographystyle{plain}

\end{document}